\documentclass[10pt, conference, review]{IEEEtran}
\IEEEoverridecommandlockouts
\usepackage{cite}
\usepackage{amsmath,amssymb,amsfonts}

\usepackage{amsthm}

\usepackage{algorithm}
\usepackage{svg}
\usepackage{algpseudocode}
\usepackage{graphicx}
\usepackage{textcomp}
\usepackage{xcolor}
\usepackage{listings}  
\usepackage{xcolor}   
\usepackage{xspace}
\usepackage{mdframed}
\usepackage{adjustbox}

\usepackage{subcaption}
\usepackage{pifont}
\usepackage{enumitem}
\usepackage{makecell}
\usepackage[nameinlink]{cleveref}
\usepackage{multirow}
\usepackage{threeparttable}
\usepackage{colortbl}
\usepackage{booktabs}
\usepackage{float}
\usepackage{todonotes}
\usepackage{balance}
\newcommand{\mpara}[1]{\smallskip\noindent{\bf #1}}

\newcommand{\multicod}{\textsc{Multi-CoD}}
\newcommand{\toolname}{\textsc{Multi-CoD}}

\usepackage{booktabs}

\usepackage{xcolor}
\definecolor{Gray}{gray}{0.9}
\usepackage{url}
\usepackage{listings}  
\usepackage{xcolor}   
\usepackage{xspace}
\usepackage{mdframed}
\usepackage{adjustbox}
\usepackage{tcolorbox}
\usepackage{subcaption}
\usepackage{pifont}
\usepackage{enumitem}
\usepackage{makecell}
\usepackage[nameinlink]{cleveref}
\usepackage{multirow}
\usepackage{threeparttable}
\usepackage{colortbl}
\newboolean{showcomments}
\setboolean{showcomments}{true}
\ifthenelse{\boolean{showcomments}}
{\newcommand{\mynote}[2]{
\fbox{\bfseries\sffamily\scriptsize#1}
{\small$\blacktriangleright$\textsf{\emph{#2}}$\blacktriangleleft$}}}
{ \newcommand{\mynote}[2]{}}

\definecolor{framecolor}{RGB}{0,0,0}
\definecolor{backcolor}{RGB}{245,245,245}
\definecolor{titlecolor}{RGB}{70,70,70}

\mdfdefinestyle{niceframe}{%
    linewidth=1pt,
    linecolor=framecolor,
    backgroundcolor=backcolor,
    roundcorner=5pt,
    innertopmargin=\baselineskip,
    frametitlefont=\bfseries\sffamily\color{titlecolor},
    frametitlealignment=\raggedright,
    frametitlebackgroundcolor=backcolor,
    shadow=true,
    shadowsize=3pt,
    shadowcolor=black!20,
}

\lstdefinelanguage{diff}{
  morecomment=[l]{\#},
  morestring=[b]",
  sensitive=true,
  morekeywords={---, +++},
  moredelim=[is][\textcolor{red}]{-}{-},
  moredelim=[is][\textcolor{blue}]{+}{+}
}

\definecolor{diffadd}{rgb}{0,0.5,0}  
\definecolor{diffrem}{rgb}{0.6,0,0}  
\definecolor{codebg}{rgb}{0.95,0.95,0.95}  

\lstdefinestyle{diff}{
  basicstyle=\ttfamily,
  backgroundcolor=\color{codebg},
  breaklines=true,
  moredelim=[l][\color{diffrem}]{-},  
  moredelim=[l][\color{diffadd}]{+},   
}
\usepackage{inconsolata}

\usepackage{graphicx}

\newtcolorbox{promptbox}[1][]{
    colback=gray!10,      
    colframe=gray!50,     
    boxrule=0.5mm,        
    arc=1mm,              
    boxsep=0mm,
    fontupper=\ttfamily\scriptsize,  
    width=0.45\textwidth,     
    title=#1,
    fonttitle=\ttfamily\footnotesize\centering,
}

\def\BibTeX{{\rm B\kern-.05em{\sc i\kern-.025em b}\kern-.08em
    T\kern-.1667em\lower.7ex\hbox{E}\kern-.125emX}}
\begin{document}

\title{Reinforcement Learning-Guided Chain-of-Draft \\for Token-Efficient Code Generation}

\author{%
  \IEEEauthorblockN{%
    \begin{tabular}{c}                                 
      Xunzhu Tang, Iyiola Emmanuel Olatunji, Tiezhu Sun, Jacques Klein, Tegawend\'e F.~Bissyand\'e \\[0.8ex]
      University of Luxembourg, Luxembourg\\
      \texttt{\{xunzhu.tang, tiezhu.sun, emmanuel.olatunji, jacques.klein, tegawende.bissyande\}@uni.lu}
    \end{tabular}%
  }%
}

\pagestyle{plain}

\maketitle



\begin{abstract}

LLMs demonstrate surface-level fluency in code generation but struggle with structured reasoning tasks requiring correctness and semantic alignment. While Chain-of-Thought (CoT) prompting enhances reasoning through intermediate steps, it suffers from verbosity and inefficiency. Chain-of-Draft (CoD) prompting offers more concise reasoning, but the stochastic nature of LLMs produces varying solution quality, making optimal selection challenging.
We propose \multicod, a reinforcement learning framework that learns to select the most promising candidate from CoD-generated solutions. Our approach uses strategy-guided prompting to encourage diverse reasoning styles and models solution selection as a contextual bandit problem. The framework optimizes interpretable features including code complexity, reasoning structure, and strategic metadata through a reward function balancing correctness, efficiency, and clarity.
Experiments on MBPP, BigCodeBench, SWE-bench Verified, and Defects4J show \multicod~outperforms and in some cases, on par with standard prompting, CoT, and CoD baselines while achieving cost and token efficiency from the user's perspective through a multi-candidate design that charges only for the selected output, reducing user billing by over 50\% and improving LLM response quality, making \multicod~more sustainable and scalable for real-world deployment. Our code is available\footnote{\url{https://anonymous.4open.science/r/MultiCoD}}.

\end{abstract}


\section{Introduction}











Large language models (LLMs) have become powerful tools for automating software engineering tasks, including code completion and program repair. Despite their fluency and generalization capabilities, LLMs often produce brittle or incorrect code, especially in tasks that require multi-step reasoning or nuanced understanding of program semantics. This limitation stems from the fact that most LLMs generate code in a single forward pass, without explicitly modeling the reasoning process that underlies correct and efficient solutions.

To address this, recent work has explored Chain-of-Thought (CoT) prompting \cite{wei2022chain}, which encourages LLMs to generate intermediate reasoning steps before producing a final answer. Originally developed for arithmetic and symbolic reasoning tasks, CoT has been adapted to code-related domains to improve correctness \cite{chen2022program, li2025structured, yin2024thinkrepair, lecodechain, paranjape2023art}. For instance, ThinkRepair~\cite{yin2024thinkrepair} adopts CoT-style prompting to guide LLMs through bug-fixing tasks, showing that structured reasoning can significantly improve repair accuracy. Similarly, Structured CoT (SCoT) \cite{li2025structured} introduces program-structure-aware reasoning steps, demonstrating that explicitly modeling control flow and logic improves code generation performance. However, one major drawback is that CoT-based approaches emphasize excessively verbose reasoning steps, which consumes a substantial number of tokens before
arriving at a final answer. This verbosity not only increases latency and token cost, but also contributes to higher energy consumption, limiting the practicality of CoT-based methods in large-scale or resource-constrained settings. Moreover, the CoT-based approach differs from human cognitive reasoning which drafts concise intermediate thoughts that capture only essential information.

Recently, the Chain-of-Draft (CoD) \cite{xu2025chain} prompting has emerged as a more efficient alternative. Inspired by how humans solve problems using minimal drafts or shorthand notes, CoD constrains each reasoning step to be succinct (e.g., $\leq$5 words), promoting clarity and modularity. While tailored for arithmetic, commonsense and symbolic reasoning tasks, we adopt and extend CoD to code generation tasks including program repair and code completion.

However, due to the inherent stochasticity of LLM outputs, naively applying CoD for code generation tasks may lead to suboptimal or inconsistent results. Even when guided by the same prompt, LLMs can produce multiple drafts that differ significantly in reasoning structure, abstraction level, or implementation strategy. This variability is further amplified when decoding parameters such as temperature are adjusted to encourage exploration. While such diversity can be beneficial, offering a rich set of candidate solutions, it also introduces a critical challenge: \textit{how to reliably identify the most promising solution among many}. This challenge is particularly critical in program repair and code completion tasks, where correctness is non-negotiable and subtle differences in logic or syntax can lead to failure.

\begin{figure}
    \centering
    \includegraphics[width=\linewidth]{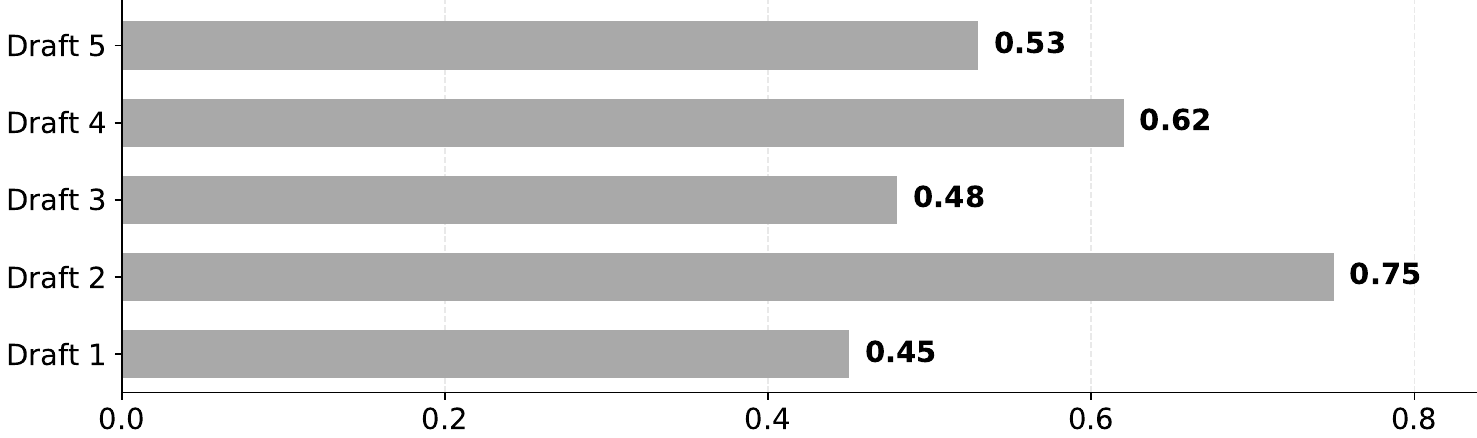}
    \caption{Comparison of BLEU scores across different solution drafts generated with identical prompting conditions. Draft 2 achieves the highest quality with a score of 0.75, while other drafts show considerable variance in performance.}
    \label{fig:bleu_comparison}
\end{figure}

\mpara{Why selecting the optimal draft matters.}
To illustrate this, consider a scenario where we generate five candidate solutions using the same CoD prompt for the \textbf{code completion task}. As shown in Figure~\ref{fig:bleu_comparison}, the quality of these solutions varies substantially despite identical generation conditions. The first draft achieves a modest BLEU score of \textbf{0.45}, while \texttt{draft\_2} reaches a significantly higher score of \textbf{0.75}. This is a 66.7\% improvement. The remaining drafts (\texttt{draft\_3}, \texttt{draft\_4}, and \texttt{draft\_5}) achieve intermediate scores of \textbf{0.48}, \textbf{0.62}, and \textbf{0.53} respectively.
This considerable variance in solution quality occurs despite all drafts being generated under identical prompting conditions, with each draft exploring different reasoning paths and implementation strategies. The observed variation highlights the inherent stochasticity of large language models and emphasizes why draft selection represents a critical yet often overlooked component in enhancing the overall effectiveness of code generation systems.
%
Therefore, the success of CoD in this domain hinges not only on generating diverse drafts, but also on selecting the optimal one. This decision must also account for both functional correctness and reasoning quality.

To address this, we propose a reinforcement learning-based framework (\multicod) that learns to select the best candidate from a set of CoD-generated solutions. By modeling solution selection as a contextual bandit problem, we train a policy agent that evaluates candidates based on interpretable features such as code complexity, reasoning structure, and strategic metadata. This approach enables us to harness the diversity of CoD while mitigating its stochasticity, leading to more robust and efficient code generation. \multicod~follows standard multi-candidate systems (e.g., GitHub Copilot generates 10 solutions but billing counts only the accepted one \footnote{https://docs.github.com/en/billing/managing-billing-for-github-copilot/about-billing-for-github-copilot}). While our approach generates multiple candidates to ensure high-quality solutions, the billing model charges only for the final selected output. This design achieves token efficiency from the user's perspective by delivering more reasoning and exploration per unit cost, while also advancing the frontier of cost-aware LLM reasoning beyond current chain-of-thought (CoT) methods. As a result, \multicod~offers a more sustainable and practical path for deploying foundation models in real-world software engineering workflows.


The contributions of this work are as follows. \begin{itemize} 
\item We identify a key limitation of Chain-of-Draft (CoD) prompting in code generation tasks: the stochastic generation of multiple drafts with varying quality, and the lack of a principled mechanism for selecting the optimal one. 
\item We propose \multicod, a reinforcement learning-based framework that learns to select the best candidate from a diverse set of CoD-generated solutions using interpretable features and a contextual bandit formulation. 
\item We introduce a strategy-guided prompting mechanism that encourages diverse reasoning styles, and a reward function that balances correctness, efficiency, and reasoning clarity. 
\item We conduct extensive experiments on four standard benchmarks (MBPP, BigCodeBench, SWE-bench, and Defects4J) across code completion and program repair, showing that \multicod~performs on par with, and in many cases outperforms existing baselines. This is achieved while significantly reducing token charges (often by over 50\%) and enabling up to 2.5× faster generation per solution, leading to lower computational cost and improved energy efficiency. These gains make \multicod~a more sustainable and scalable approach for deploying LLMs in real-world software engineering tasks.

\end{itemize}


\section{The \multicod~Framework}
\label{sec:method}

\begin{figure*}
    \centering
    \includegraphics[width=\linewidth]{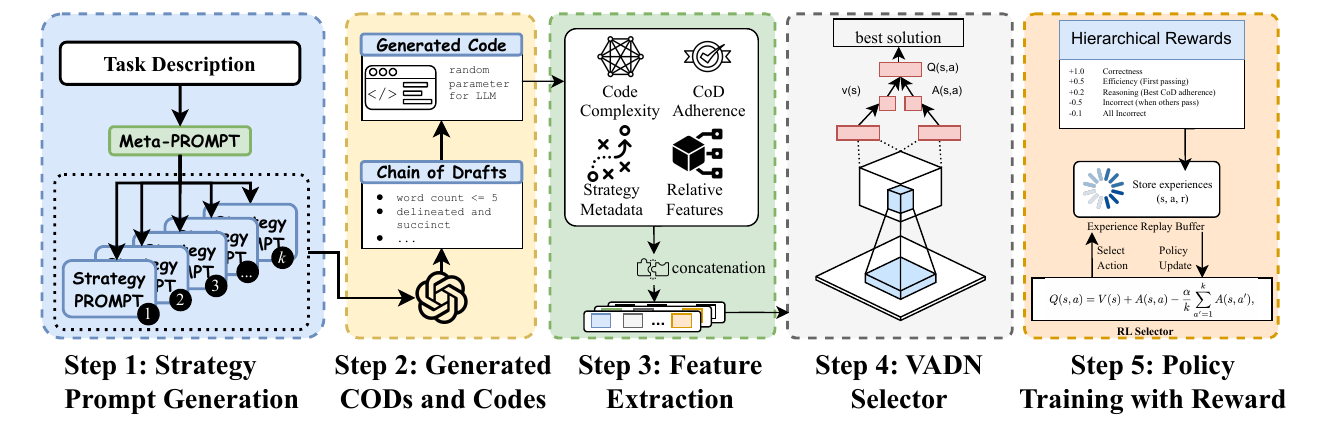}
    \caption{Overview of the \multicod~framework. We generate $k$ CoD-guided solutions with diverse strategies and select the best using a learned reinforcement learning selector.}
    \label{fig:pipeline}
\end{figure*}

We present \multicod, a reinforcement learning-based code generation framework that selects high-quality solutions from a set of diverse Chain-of-Draft (CoD) candidates. As shown in Figure~\ref{fig:pipeline}, the framework consists of five main stages: (1) strategy-guided prompt generation, (2) CoD-constrained solution synthesis, (3) feature encoding, (4) RL-based candidate selection, and (5) policy training with reward shaping.

\subsection{Problem Formulation}

Given a programming task consisting of a description $t$ and a code signature $h$, our objective is to produce a correct and well-structured solution by first generating $k$ candidates $\{c_1, \dots, c_k\}$ and selecting the most promising one via a learned policy $\pi$. We aim to learn a policy that maximizes the expected reward across tasks:

\begin{equation}
\pi^* = \arg\max_\pi \mathbb{E}_{t} \left[ r(s, \pi(s)) \right]
\end{equation}

Here, $s$ is the state encoding of all candidate solutions, $\pi(s)$ is the selected index, and $r(\cdot, \cdot)$ is the scalar reward.

\subsection{Step 1: Strategy-Guided Prompt Generation}

We systematically produce diverse CoD prompts, each embodying distinct problem-solving strategies. Given a task defined by $(t, h)$, we formulate the prompt-generation process as:
\begin{equation}
\mathcal{P}(t, h) \rightarrow \{p_i\}_{i=1}^{k}, \quad \text{s.t.} \quad \text{diversity}(\{p_i\}) \geq \tau,
\end{equation}
where $k = 5$ prompts are generated and $\tau$ ensures strategic variation.

To quantify strategic diversity, we compute pairwise cosine distances between prompt embeddings:
\begin{equation}
\small
\text{diversity}(\{p_i\}) = \frac{2}{k(k-1)} \sum_{i=1}^{k-1}\sum_{j=i+1}^{k}\left[1 - \frac{E(p_i)\cdot E(p_j)}{\|E(p_i)\|\|E(p_j)\|}\right],
\end{equation}
where $E(p_i)$ denotes the embedding vector for prompt $p_i$.

Each prompt $p_i$ contains five structured components: (1) \textbf{strategy name} summarizing the strategic perspective, (2) \textbf{strategic focus} identifying the optimization target (performance, readability, robustness), (3) \textbf{instruction} providing actionable guidance, (4) \textbf{key priorities} enumerating implementation priorities, and (5) \textbf{fully realized prompt text} for the final solution-generation (based on front 4 steps) step.

\subsection{Step 2: CoD-Constrained Solution Synthesis}

We synthesize candidate solutions through structured decoding, constrained by the CoD methodology requiring each reasoning step to be $\leq$5 words.

\mpara{Draft Generation Phase.} 
The model generates reasoning steps $D_i = \{d_{i,j}\}_{j=1}^{m}$ where the CoD constraint is enforced:
\begin{equation}
\text{valid}(d_{i,j}) = \mathbb{I}\left[\,\text{word\_count}(d_{i,j}) \leq 5\,\right],
\quad \forall j \in \{1, \dots, m\},
\end{equation}
Steps violating this constraint trigger regeneration until satisfied.

\mpara{Code Generation Phase.}
Final executable code $c_i$ is synthesized conditioned on the constrained reasoning steps $D_i$:
\begin{equation}
P(c_i \mid D_i, p_i) = \prod_{t=1}^{T_i} P\left(c_{i,t} \mid c_{i,<t}, D_i, p_i\right),
\end{equation}
where $c_{i,t}$ represents the $t^{th}$ token in the code.

To encourage diversity, we vary the decoding temperature parameter $\tau_i \in \{0.4, 0.5, 0.6, 0.7, 0.8\}$ across candidates.

\subsection{Step 3: Comprehensive Feature Extraction}

With multiple candidate solutions synthesized under diverse strategic constraints, we systematically extract comprehensive, interpretable features from each candidate. These features form essential input for the subsequent reinforcement learning-based selection policy.

Formally, for a given candidate solution $c_i$, draft $D_i$, and prompt $p_i$, we define a structured feature extraction function:
\begin{equation}
F(c_i, D_i, p_i) \rightarrow f_i \in \mathbb{R}^{m}.
\end{equation}
where $m=26$ represents the total number of features that capture key aspects of code structure, reasoning behavior, and prompt metadata (see ~\Cref{tab:features} for details). Note that the selection of the features is driven by the need to balance expressiveness, efficiency, and interpretability. Therefore, we exclude semantic representations such as AST embeddings, control-flow graphs, and data-dependency graphs, as these introduce 5–10x higher runtime cost and significantly reduce interpretability.


\begin{table}[htbp]
\centering
\footnotesize
\caption{Extracted features for candidate solutions.}
\label{tab:features}
\resizebox{0.9\linewidth}{!}{
\begin{tabular}{@{}c|l|l@{}}
\toprule
 & \textbf{Feature Name} & \textbf{Definition} \\ \midrule

\multirow{10}{*}{\rotatebox[origin=c]{90}{\textbf{Code Complexity}}} 
& Char count (log) & Logarithmic count of characters in solution code \\
& Line count (log) & Logarithmic count of code lines \\
& Function count (log) & Logarithmic number of functions defined \\
& Loop count (log) & Logarithmic count of loops (\texttt{for}, \texttt{while}) \\
& Conditional count (log) & Logarithmic count of conditionals (\texttt{if}, \texttt{elif}) \\
& Try-catch count (log) & Logarithmic count of error-handling blocks \\
& Import count (log) & Logarithmic count of imported modules \\
& Class count (log) & Logarithmic number of defined classes \\
& Comment count (log) & Logarithmic count of code comments \\
& Avg line length & Average number of characters per line \\ \midrule

\multirow{6}{*}{\rotatebox[origin=c]{90}{\textbf{Chain-of-Draft}}} 
& Adherence rate & Ratio of valid CoD steps ($\leq$ 5 words) to total steps \\
& Total draft steps (log) & Logarithmic total number of CoD reasoning steps \\
& Avg words per step & Average word count per reasoning step \\
& Min words per step & Minimum word count among steps \\
& Max words per step & Maximum word count among steps \\
& Std dev words per step & Standard deviation of word counts across steps \\ \midrule

\multirow{4}{*}{\rotatebox[origin=c]{90}{\textbf{Strategy}}} 
& Strategy index & Identifier index of prompt strategy used \\
& Temperature ($\tau_i$) & Temperature setting during generation \\
& Time-focused & 1 if strategy mentions speed/performance optimization, else 0 \\
& Space-focused & 1 if strategy mentions memory/storage optimization, else 0 \\ \midrule

\multirow{6}{*}{\rotatebox[origin=c]{90}{\textbf{Relative}}} 
& Char count ratio & Character count normalized by longest candidate \\
& Line count ratio & Line count normalized by longest candidate \\
& Draft steps ratio & Draft steps count normalized by maximum among candidates \\
& Rank by length & Normalized rank based on code length \\
& Rank by adherence & Normalized rank based on CoD adherence rate \\
& Is shortest & Binary indicator if candidate is shortest among all \\ \bottomrule
\end{tabular}}
\end{table}

Features are normalized and concatenated as:
\begin{equation}
f_i = [\,f_i^{\text{code}};\; f_i^{\text{cod}};\; f_i^{\text{meta}};\; f_i^{\text{rel}}\,] \in \mathbb{R}^{26}.
\end{equation}

These features form essential input for the subsequent reinforcement learning-based selection policy.

\subsection{Step 4: Reinforcement Learning-Based Solution Selection}

We frame selection as a contextual bandit problem. Given state representation $s \in \mathbb{R}^{k \times 26}$, we select the optimal candidate:
\begin{equation}
a^* = \arg\max_{a \in \{1,\dots,k\}} Q(s, a),
\end{equation}
where $Q(s, a)$ is the action-value function.

\mpara{Value-Advantage Decomposition Network (VADN).}
We propose VADN with adaptive scaling, layer normalization, and residual connections. The architecture decomposes $Q(s,a)$ as:
\begin{equation}
Q(s,a) = V(s) + A(s,a) - \frac{\alpha}{k}\sum_{a'=1}^{k} A(s,a'),
\end{equation}
where $V(s)$ is the state-value function, $A(s,a)$ is the advantage function, and $\alpha$ is a learnable scaling parameter.

The architecture includes:

\paragraph{Feature Processing Layer} 
\begin{equation}
h_s = \text{LayerNorm}(\text{ReLU}(W_f \cdot s + b_f)),
\end{equation}

\paragraph{Value Stream} with residual connections:
\begin{equation}
V(s) = W_v \cdot (\text{ReLU}(h_s) + h_s) + b_v,
\end{equation}

\paragraph{Advantage Stream} with two-layer architecture:
\begin{equation}
\begin{split}
h_a &= \text{ReLU}(W_{a1} \cdot h_s + b_{a1}), \\
A(s,a) &= W_{a2} \cdot h_a + b_{a2}.
\end{split}
\end{equation}

Training minimizes the composite loss:
\begin{equation}
\mathcal{L}(\theta) = \mathbb{E}_{(s,a,r)}\left[\left(r + \gamma\max_{a'}Q_{\theta^-}(s,a') - Q_\theta(s,a)\right)^2\right] + \lambda \|\theta\|_2^2,
\end{equation}
using prioritized experience replay for efficient learning.

\subsection{Step 5: Reward Design and VADN Policy Training}

We define a hierarchical reward structure prioritizing correctness, efficiency, and reasoning quality:
\begin{equation}
r(s,a) = 
\begin{cases}
    1.0 + 0.5 \cdot \mathbb{I}[\text{first\_pass}(c_a)] + \\
    \quad 0.2 \cdot \mathbb{I}[\text{best\_adh}(D_a)], & \text{if } c_a \text{ passes tests}, \\
    -0.5, & \text{if } c_a \text{ fails, others pass}, \\
    -0.1, & \text{if all } c_i \text{ fail},
\end{cases}
\end{equation}
where $\mathbb{I}[\text{first\_pass}(c_a)]$ rewards efficiency and $\mathbb{I}[\text{best\_adh}(D_a)]$ encourages interpretable reasoning.

Training uses prioritized experience replay with experiences weighted by temporal difference error. We employ adaptive epsilon-greedy exploration with dynamic decay and separate optimization for the scaling parameter $\alpha$.

\subsection{Inference Process}

During inference, we generate $k$ CoD candidates, extract features, and select using the trained VADN policy:
\begin{equation}
a^* = \arg\max_{a \in \{1,\dots,k\}} \left[V(s) + A(s,a) - \frac{\alpha}{k}\sum_{a'=1}^{k} A(s,a')\right].
\end{equation}

Only the selected solution is evaluated, yielding a 5× reduction in execution cost while preserving solution quality.

\section{Experimental Setup}
\label{sec:setup}

In this section, we describe our experimental methodology for evaluating \toolname, including the benchmarks, evaluation metrics, baselines, and experimental settings.

\subsection{Benchmarks}

We evaluate our approach using four diverse benchmarks summarized in Table~\ref{all-benches}.

\begin{table}[h]
\centering
\caption{Overview of Benchmark Tasks and Dataset Statistics}
\resizebox{\linewidth}{!}{
\vspace{-0.2cm}
\begin{tabular}{lccc}
\toprule
\textbf{Benchmark} & \textbf{Task Type} & \textbf{Total Samples} \\
\midrule
MBPP & Code Generation & 974   \\
SWE-bench Verified & Program Repair & 500 \\
Defects4J & Program Repair & 835  \\
BigCodeBench & Code Completion & 1,140  \\
\bottomrule
\end{tabular}
}
\label{all-benches}
\vspace{-0.3cm}
\end{table}


\begin{itemize}[leftmargin=*]
    \item \textbf{BigCodeBench}~\cite{lai2023bigcodebench}: A comprehensive benchmark containing a diverse set of coding problems across multiple programming languages and difficulty levels. It provides a robust test of general code generation and repair capabilities. We use the ``hard'' variant and ``complete'' prompt style.
    
    \item \textbf{MBPP (Mostly Basic Programming Problems)}~\cite{austin2021program}: A collection of 964 programming problems designed to evaluate basic programming skills. These problems are typically shorter and more focused than those in other benchmarks.
    
    \item \textbf{SweBench-verified}~\cite{jimenez2023swebench}: A benchmark specifically designed for software engineering tasks, with verified solutions that include unit tests to validate functional correctness. This benchmark focuses on real-world programming scenarios.
    
    \item \textbf{Defects4J}~\cite{just2014defects4j}: A database of real bugs from open-source Java projects, providing a challenging test of program repair capabilities on production-level code. Each bug comes with a corresponding test suite that can be used to validate repairs.
\end{itemize}

\subsection{Evaluation Metrics}

We employ the following metrics to evaluate 
performance:

\begin{itemize}[leftmargin=*]
    \item \textbf{Pass@1~\cite{chen2021evaluating}}. The percentage of problems for which the model's first generated solution passes all test cases. This metric evaluates the model's ability to produce correct repairs on the first attempt.


\item \textbf{Accuracy~\cite{austin2021program}}. The percentage of programming problems for which the model generates code that passes all test cases. This metric is commonly used on datasets like MBPP to measure the model's ability to correctly implement a solution based on a natural language problem description.

    \item \textbf{Compilation Rate (CR)~\cite{just2014defects4j}}:  The percentage of generated or modified code that successfully compiles without errors.

    \item \textbf{BLEU~\cite{papineni2002bleu}}. This metric calculates the geometric mean of modified n-gram precisions, penalized by a brevity penalty, to evaluate the quality of machine-generated translations against one or more reference translations

\item \textbf{Resolved~\cite{jimenez2023swebench}}. The percentage of task instances where the generated patch applies successfully and passes all test cases. A solution is considered to have ``resolved'' the issue when it can be cleanly applied to the codebase and satisfies all the specified test requirements.

    
    
\end{itemize}


    

\subsection{Models}

We compare the following models and variants in our evaluation:

\begin{itemize}[leftmargin=*]
    \item \textbf{Closed-Source Model}. We utilize four popular closed-source LLMs as our foundation models. Specifically, Claude-Sonnet3.5 and 3.7~\cite{Anthropic2025}, GPT-4o and GPT-o1~\cite{openai2023gpt4}, which are state-of-the-art models demonstrating strong performance in coding and reasoning.
    
    \item \textbf{Base Open-Source Model}. We also consider Qwen2.5-Coder-32B-Instruct and Qwen3-30B-A3B~\cite{hui2024qwen2} as representative open-source models.
\end{itemize}

\subsection{Prompt Strategy}

We utilize three prompt strategies as baselines: standard prompting, CoT, and CoD.\\
\noindent{\bf Standard prompting.} We use standard few-shot prompting~\cite{fewshot}, where the model is given input-output pairs as in-context examples. LLMs are asked to directly return the final answer, without any reasoning or explanation. 

\noindent{\bf Chain-of-Thought.}
We follow the exact few-shot examples provided in the appendix of the CoT paper with the exception of having the final answer after four hashtags ({\small \#\#\#\#}) for a more stable answer extraction. 

\noindent{\bf Chain-of-Draft.} CoD is a CoT similar strategy, thinking step by step. However, the model is asked to limit each reasoning step to five words at most and keep the key point of the reasoning plan.

\begin{figure}
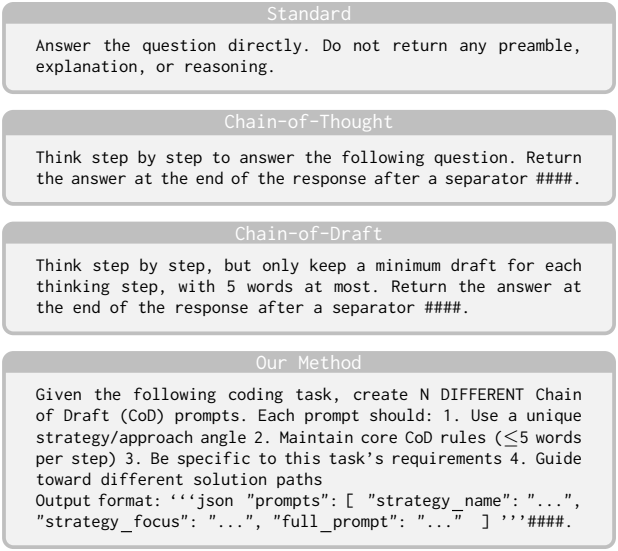

    \centering
\begin{center}
\centering
\begin{promptbox}[Standard]
Answer the question directly.
Do not return any preamble, explanation, or reasoning.
\end{promptbox}
\begin{promptbox}[Chain-of-Thought]
Think step by step to answer the following question.
Return the answer at the end of the response after a separator \#\#\#\#.
\end{promptbox}
\begin{promptbox}[Chain-of-Draft]
Think step by step, but only keep a minimum draft for each thinking step, with 5 words at most.
Return the answer at the end of the response after a separator \#\#\#\#.
\end{promptbox}
\begin{promptbox}[Our Method]
Given the following coding task, create \textit{N} DIFFERENT Chain of Draft (CoD) prompts. Each prompt should:
1. Use a unique strategy/approach angle
2. Maintain core CoD rules ($\leq$5 words per step)
3. Be specific to this task's requirements
4. Guide toward different solution paths

Output format:
```json
{
  "prompts": [
    {
      "strategy\_name": "...",
      "strategy\_focus": "...",
      "full\_prompt": "..."
    }
  ]
}'''\#\#\#\#.
\end{promptbox}
\end{center}
    \caption{Prompts in different methods.}
    \label{fig:prompts}
\end{figure}

The complete system prompt for each prompting strategy is shown in Figure~\ref{fig:prompts}.


\subsection{Experimental Procedure}
Our experimental procedure consists of the following steps:
\begin{enumerate}[leftmargin=*]
    \item \textbf{CoD Prompt Generation}: Given the meta prompt, task, and LLM, we query LLM to generate \textit{k} CoD prompts.
    \item \textbf{Code Generation}: We query LLM with each CoD to generate codes step by step, under the requirement: ``Maintain core CoD rules ($\leq$5 words per step) and Be specific to this task's requirements''.
    
    \item \textbf{Code Encoding}: We consider 26 features in 4 aspects to represent the characteristics of generated codes. These four aspects are: code complexity, CoD Adherence, Strategy Metadata, and Relative Features. Then, we will use the initial embedding layer of Pytorch~\cite{imambi2021pytorch} to encode these features and concatenate them into the final vector. Here, we will have \textit{k} vectors for \textit{k} snippets of code, respectively. Additionally, for each generated snippet of code, we concatenate the different features.
    
    \item \textbf{VADN Solution Selector}: We implement our Value-Advantage Decomposition Network (VADN) using a two-stream architecture with shared feature processing layers. The network consists of a feature processing layer with LayerNorm, a value stream with residual connections, and a two-layer advantage stream. The model takes the concatenated feature vectors from all candidates as input and outputs Q-values for each candidate. We initialize the adaptive scaling parameter $\alpha$ to 1.0 and set the network dimensions to [26, 64, 32] for the shared layers, with output dimensions of 1 for the value stream and \textit{N} for the advantage stream.
    
    \item \textbf{Reward Model design and VADN policy training}: We define a hierarchical reward function that prioritizes correctness (+1.0), efficiency (+0.5 for first passing solution), and reasoning quality (+0.2 for best CoD adherence). We implement prioritized experience replay with a buffer size of 10,000 and train the VADN using a learning rate of 0.001 for network parameters and 0.01 for the adaptive scaling parameter. We use an epsilon-greedy policy with adaptive decay based on performance, starting with $\epsilon=0.3$ and decaying to a minimum of 0.05. Training proceeds for 200 epochs with batch size 32, with the target network updated every 10 epochs.
    
    \item \textbf{Inference}: During evaluation, we generate 5 diverse CoD solutions for each test problem, extract their feature vectors, and select a single candidate using the trained VADN policy. Only the selected solution is evaluated against test cases. We record pass rates, execution times, and CoD adherence metrics for the selected solutions. For comparative analysis, we also record the performance of a uniform random selector and the oracle (best possible) selector as baselines.
\end{enumerate}


\subsection{Implementation Details}

We implement \multicod~using a mix of closed- and open-source LLMs. For closed-source models, we use Claude-3-7-Sonnet, Claude-3-5-Sonnet, GPT-4o, and GPT-o1 via API access on a 32-core AMD EPYC server with 16 parallel processes. For open-source models, we deploy Qwen3-30B-A3B and Qwen2.5-Coder-32B-Instruct on 8× NVIDIA H100 GPUs (640GB total VRAM).

The VADN selector is implemented in PyTorch with a three-layer architecture comprising a shared feature encoder, a value stream, and an advantage stream. 
The selector extracts 26 features across code complexity (10 log-scaled metrics), CoD adherence (6 metrics including word counts), strategy metadata (4 indicators), and relative comparisons (6 normalized features). For the training, we use 200 BigCodeBench tasks (17.5\% of dataset) with a three-layer architecture: shared encoder (26$\rightarrow$128), value stream (128$\rightarrow$64$\rightarrow$1), and advantage stream (128$\rightarrow$64$\rightarrow$5). We use the following hyperparameters: learning rate $10^{-4}$, $\epsilon$-greedy exploration starting at 0.3 with 0.995 decay, experience replay buffer of 10,000, batch size 32, and 300 training epochs. The values of the hierarchical reward structure were selected via a random search.


\subsection{VADN Training}

We train the Value-Advantage Decomposition Network (VADN) on a subset of 200 tasks from the BigCodeBench dataset (approximately 17.5\% of the full set). For each task, $k=5$ candidate solutions are generated using CoD-constrained prompting with pre-trained foundation models (Claude, GPT, and Qwen variants). These models are used as-is, without any fine-tuning. All generated solutions are evaluated against ground-truth test cases to compute rewards, which guide the policy learning process.

During training, VADN learns to identify high-quality solutions based on interpretable features such as code complexity, reasoning structure, and strategic metadata. Once trained, the VADN selector is applied without modification across all four benchmarks (BigCodeBench, MBPP, SWE-bench Verified, and Defects4J). Despite being trained on a single dataset, the model generalizes effectively, demonstrating its ability to capture transferable patterns in code quality and reasoning across diverse tasks and domains.

\subsection{Token Usage and Generation Time Cost Computation}
\label{sec:token-time-gen-compute}
We measure token usage and generation time cost in two stages: (i) \textit{strategy generation}, where $k$ CoD prompts are produced, and (ii) \textit{solution generation}, where $k$ code solutions are generated for the corresponding strategies. 
For \multicod(best), token usage and generation time cost is the sum of all strategy generation and a single selected solution. In contrast, \multicod(all) accounts for all strategy generation and all $k$ solutions. Unless otherwise stated, references to \multicod~in token and time-related discussions default to \multicod(best).
Note that since strategy and solution can be parallelized, the generation time cost for \multicod(all) can be reduced to that of \multicod(best).

\section{Experimental Results}
\label{sec:results}

In this section, we present a comprehensive evaluation of \multicod{}~ across multiple program repair and code generation benchmarks. We present results related to 
quantitative performance analysis, ablation studies, case studies, and generalization experiments.

\subsection{Performance Comparison}
We apply \multicod{}~ on each benchmark and compute performance based on metrics associated with the given benchmark. We further report, for comparison, the performance metrics of other state-of-the-art or baseline models for each benchmark.

\subsubsection{BigCodeBench}

\begin{figure}[ht]
    \centering
    \includegraphics[width=0.95\linewidth]{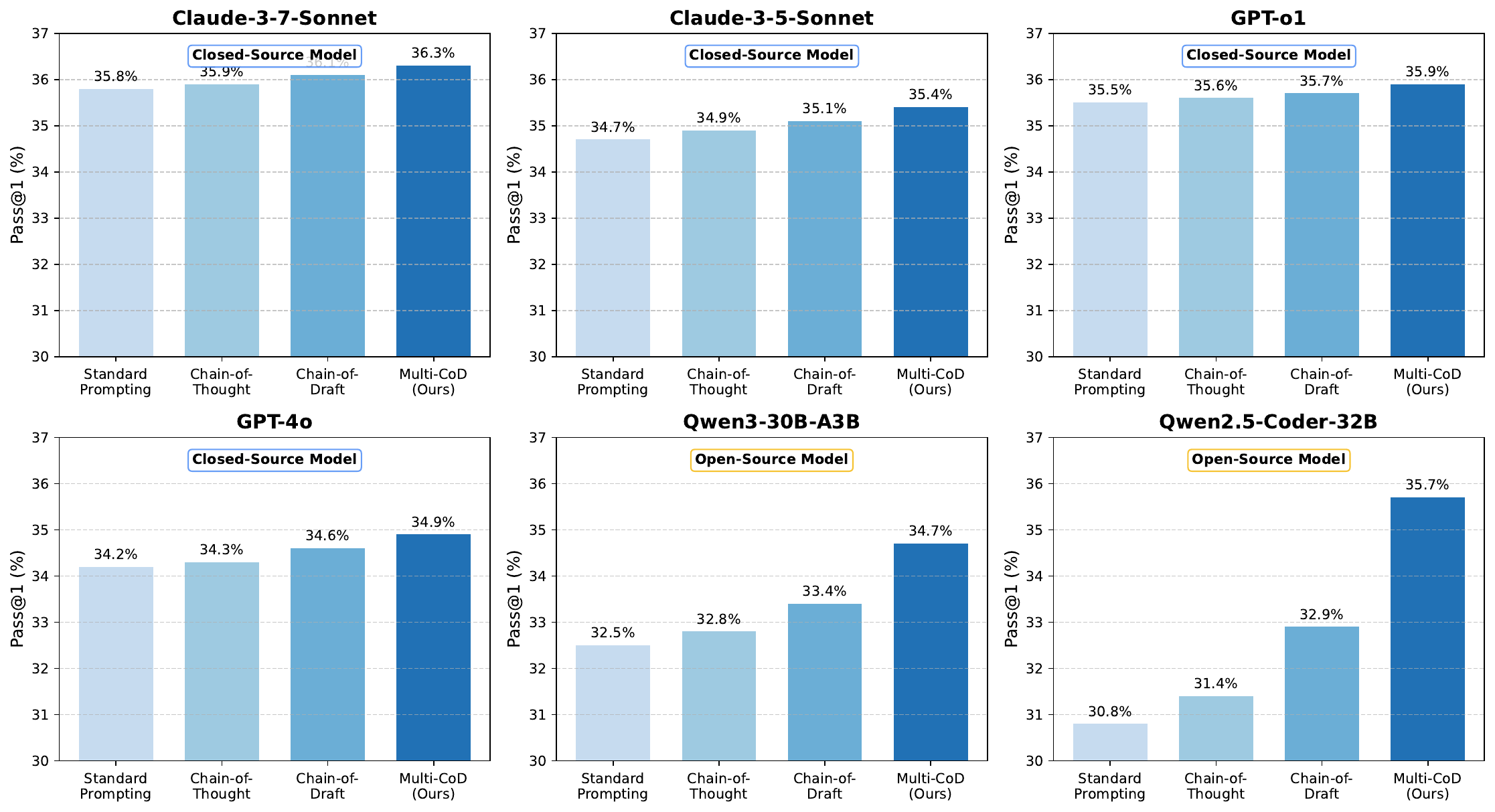}
    \caption{Performance comparison of prompting strategies across different foundation models on BigCodeBench (Pass@1). Each axis represents a different prompting strategy, with values increasing outward. Solid lines represent closed-source models, while dashed lines represent open-source models.}
    \label{fig:bigcodebench_radar}
\end{figure}

On BigCodeBench, we evaluate different prompting strategies across closed-source and open-source foundation models. Figure \ref{fig:bigcodebench_radar} visualizes these results, clearly demonstrating how our \multicod{}~framework performs across all model configurations.

The visualization reveals several key insights: (1) \multicod{}~with Claude-3-7-Sonnet achieves a Pass@1 rate of 36.3\%, establishing a new state-of-the-art that outperforms standard Claude-3-7-Sonnet by 0.5\%; (2) Chain-of-Draft consistently outperforms both Standard Prompting and Chain-of-Thought across all models; (3) Open-source models show dramatically larger gains from our approach, with Qwen2.5-Coder-32B-Instruct improving by 4.9\% to reach 35.7\% - effectively closing the gap with several closed-source alternatives.

Table \ref{tab:bigcodebench_multicod} highlights the performance of our \multicod{}~approach compared to the previous state-of-the-art.

\begin{table}[!h]
    \centering
    \caption{BigCodeBench Pass@1 Results, highlighting \multicod{}'s improvements}
    \begin{tabular}{l|c}
        \toprule
        \textbf{Model} & \textbf{Pass@1} \\
        \midrule
        \rowcolor{Gray} \textbf{\multicod{}~+ Claude-3-7-Sonnet (Ours)} & \textbf{36.3} \\
        Claude-3-7-Sonnet (Standard) & 35.8 \\
        GPT-o1 & 35.5 \\
        \multicod{}~+ Qwen2.5-Coder-32B & 35.7 \\
        GPT-4o & 34.2 \\
        Qwen2.5-Coder-32B (Standard) & 30.8 \\
        \bottomrule
    \end{tabular}
    \label{tab:bigcodebench_multicod}
\end{table}

\subsubsection{MBPP}
\begin{figure}[ht]
    \centering
    \includegraphics[width=\linewidth]{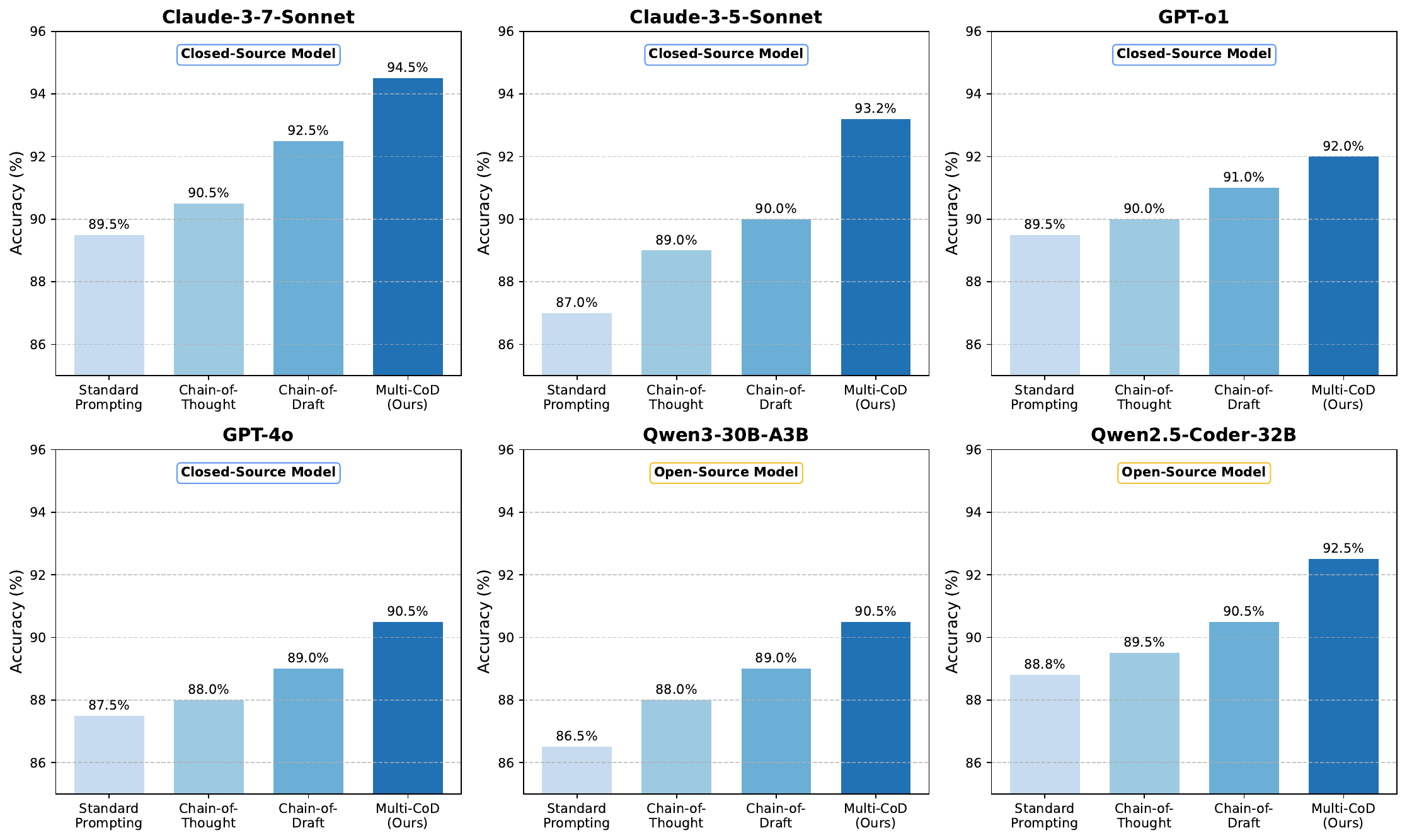}
    \caption{MBPP benchmark performance by model and prompting strategy. Each panel shows a different foundation model's accuracy progression across four prompting methods: Standard Prompting, Chain-of-Thought, Chain-of-Draft, and \multicod{}. Blue shades represent increasingly sophisticated prompting strategies.}
    \label{fig:mbpp_model_charts}
\end{figure}

On the MBPP benchmark, we analyze the impact of different prompting strategies across both closed-source and open-source foundation models. Figure \ref{fig:mbpp_model_charts} presents individual performance charts for each model, clearly illustrating the progressive improvements achieved through our sequence of prompting methodologies.

The results reveal a consistent pattern of improvement across all models. Our \multicod{}~framework with Claude-3-7-Sonnet achieves 94.5\% accuracy, establishing a new state-of-the-art that surpasses previous specialized approaches like QualityFlow (94.2\%) and o1-mini + MapCoder (93.2\%). Notably, the performance gains are particularly pronounced for open-source models, with Qwen2.5-Coder-32B improving from 88.8\% with standard prompting to 92.5\% with \multicod{}~(+3.7\%), effectively outperforming several standard closed-source configurations.

Several key insights emerge from our analysis: (1) Each progressive prompting strategy contributes incremental improvements, with the largest gains typically observed between Chain-of-Draft and \multicod{}; (2) The closed-source Claude-3-7-Sonnet consistently demonstrates the highest performance ceiling, but all models benefit significantly from our approach; (3) The performance gap between open and closed-source models substantially narrows with advanced prompting strategies, suggesting that sophisticated prompting can partially compensate for differences in model scale or training.

\begin{table}[!h]
    \centering
    \caption{MBPP Accuracy Results, highlighting \multicod{}'s improvements over existing approaches.}
    \resizebox{0.85\linewidth}{!}{
    \begin{tabular}{l|c}
        \toprule
        \textbf{Model} & \textbf{Accuracy} \\
        \midrule
        \rowcolor{Gray} \textbf{\multicod{}~+ Claude-3-7-Sonnet (Ours)} & \textbf{94.5} \\
        QualityFlow (Sonnet-3.5) & 94.2\\
        \multicod{}~+ Claude-3-5-Sonnet (Ours) & 93.2 \\
        o1-mini + MapCoder (Hamming.ai) & 93.2\\
        \multicod{}~+ Qwen2.5-Coder-32B (Ours) & 92.5\\
        \multicod{}~+ GPT-o1 (Ours) & 92.0 \\
        Chain-of-Draft + Claude-3-7-Sonnet & 92.5 \\
        \multicod{}~+ Qwen3-30B-A3B (Ours) & 90.5 \\
        Claude-3-7-Sonnet (Standard) & 89.5\\
        Qwen2.5-Coder-32B (Standard) & 88.8\\
        GPT-4o (Standard) & 87.5\\
        \bottomrule
    \end{tabular}}
    \label{tab:mbpp}
\end{table}

\subsubsection{SWE-bench Verified}
\begin{figure}[ht]
    \centering
    \includegraphics[width=\linewidth]{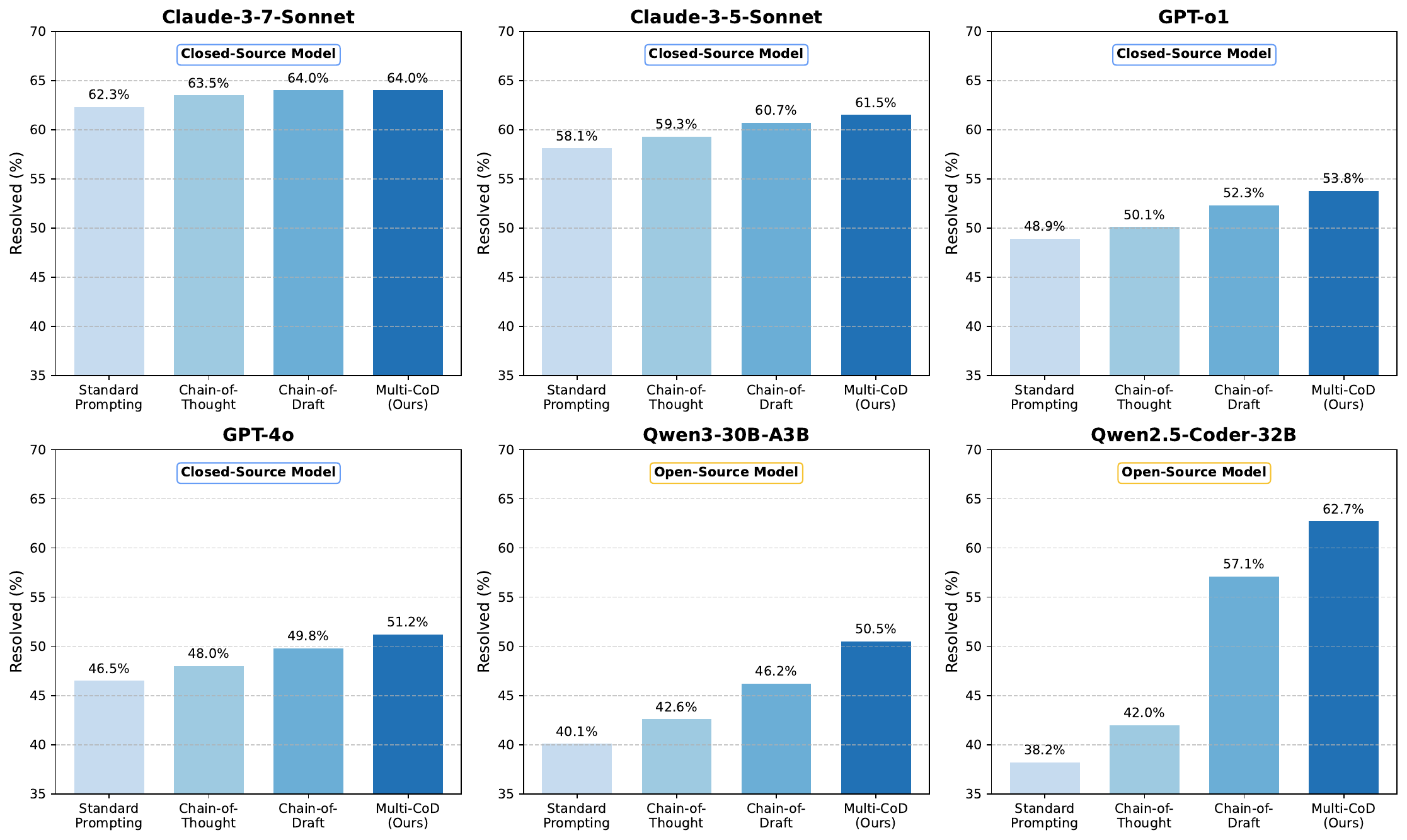}
    \caption{SWE-bench Verified performance by model and prompting strategy. Each panel shows a different foundation model's resolution rate progression across four prompting methods: Standard Prompting, Chain-of-Thought, Chain-of-Draft, and \multicod{}. The most dramatic improvement is observed in Qwen2.5-Coder-32B.}
    \label{fig:swebench_model_charts}
\end{figure}

On the SWE-bench Verified benchmark, our approach delivers particularly dramatic improvements across different foundation models. Figure \ref{fig:swebench_model_charts} presents individual performance charts for each model, highlighting the substantial gains achieved through our sequence of prompting methodologies.

The most striking observation is the exceptional improvement in the open-source Qwen2.5-Coder-32B model, which achieves a resolution rate of 62.7\% with our \multicod{}~framework—a substantial 24.5\% absolute improvement over its standard prompting performance (38.2\%). This dramatic gain enables it to match or exceed the performance of significantly more powerful closed-source models including Claude-3-7-Sonnet in its default configuration (62.3\%) and substantially outperform other models like GPT-o1 (48.9\%) and o3-mini (49.3\%).

Our results reveal a consistent pattern across models, with Chain-of-Draft showing substantial gains over previous methods, and \multicod{}~further enhancing performance through its strategic diversity and reinforcement learning-based selection. Notably, the performance gap between open and closed-source models narrows considerably with sophisticated prompting strategies, suggesting that prompt engineering can partially compensate for differences in model scale and training data.

\begin{table}[!h]
    \centering
    \caption{SWE-bench Verified Results}
    \resizebox{0.9\linewidth}{!}{
    \begin{tabular}{l|c}
        \toprule
        \textbf{Model} & \textbf{Resolved (\%)} \\
        \midrule
        Claude 3.7 Sonnet (with scaffold) & 70.3\\
        \rowcolor{Gray} \multicod{}~+ Claude-3-7-Sonnet (Ours) & 64.0\\
        \multicod{}~+ Qwen2.5-Coder-32B (Ours) & 62.7\\
        Claude 3.7 Sonnet (default) & 62.3\\
        \multicod{}~+ Claude-3-5-Sonnet (Ours) & 61.5 \\
        \multicod{}~+ GPT-o1 (Ours) & 53.8 \\
        \multicod{}~+ Qwen3-30B-A3B (Ours) & 50.5 \\
        o3-mini & 49.3\\
        o1 & 48.9\\
        Nebius AI Qwen 2.5 72B + Llama 3.1 70B & 40.6\\
        Qwen2.5-Coder-32B (Standard) & 38.2\\
        \bottomrule
    \end{tabular}}

    \label{tab:swebench}
\end{table}

These results demonstrate that \multicod{}~is particularly effective for complex software engineering tasks that require structured reasoning and strategic diversity. The framework's ability to dramatically improve open-source model performance positions it as a cost-effective alternative to proprietary systems for real-world software development tasks.

\subsubsection{Defects4J}
\begin{figure}[ht]
    \centering
    \includegraphics[width=\linewidth]{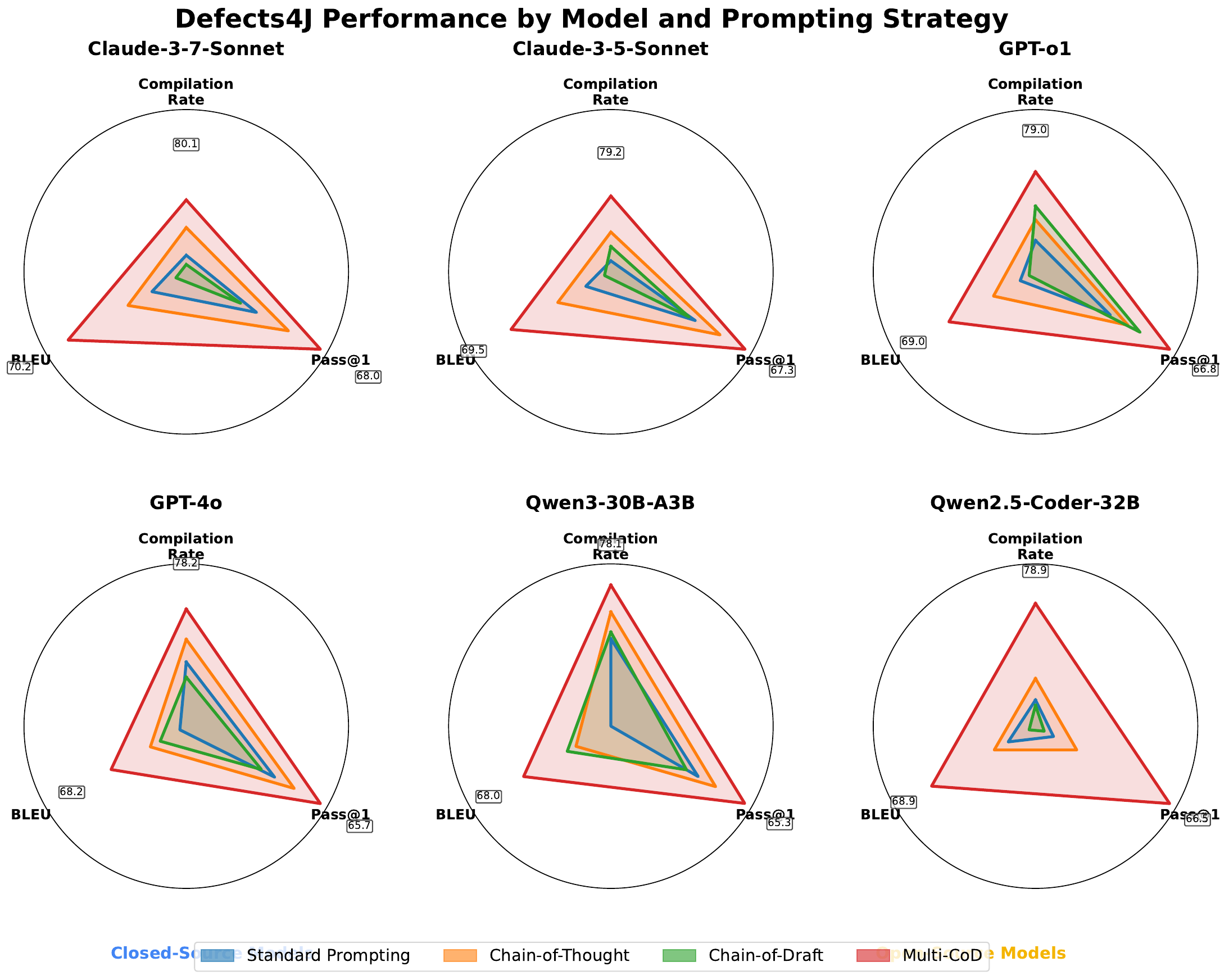}
    \caption{Defects4J performance visualization across three metrics (Compilation Rate, Pass@1, and BLEU) for each foundation model. Each radar chart uses normalized axes to highlight the differences between prompting strategies. Actual metric values are shown as labels.}
    \label{fig:defects4j_radar}
\end{figure}

For Defects4J, we analyze performance across three critical metrics: Compilation Rate (CR), Pass@1, and BLEU score. Figure \ref{fig:defects4j_radar} presents individualized radar charts for each foundation model, highlighting the subtle but important performance differences across prompting strategies.

Interestingly, Defects4J exhibits a more nuanced pattern compared to other benchmarks. While Chain-of-Draft shows consistent improvements for most models on most benchmarks, here we observe occasional performance dips, particularly for Qwen2.5-Coder-32B where Chain-of-Draft underperforms Standard Prompting on all metrics. This suggests that the highly constrained reasoning approach of Chain-of-Draft may occasionally interfere with effective program repair, which often requires flexible contextual understanding.

Nevertheless, our \multicod{}~framework consistently recovers from these dips and delivers the strongest performance across all models and metrics by intelligently selecting from diverse candidate solutions. For Qwen2.5-Coder-32B, \multicod{}~improves Compilation Rate from 77.1\% to 78.9\%, Pass@1 from 64.0\% to 66.5\%, and BLEU score from 67.8\% to 68.9\%. Even more impressively, \multicod{}~enhances the already strong performance of Claude-3-7-Sonnet, increasing its Compilation Rate from 79.5\% to 80.1\%, Pass@1 from 67.2\% to 68.0\%, and BLEU score from 69.5\% to 70.2\%.

\begin{table}[!h]
    \centering
    \caption{Defects4J Results}
    \resizebox{\linewidth}{!}{
    \begin{tabular}{l|c|c|c}
        \toprule
        \textbf{Model} & \textbf{Compilation Rate (CR) (\%)} & \textbf{Pass@1} & \textbf{BLEU} \\
        \midrule
        \rowcolor{Gray}\multicod{}~+ Claude-3-7-Sonnet & \textbf{80.1} & \textbf{68.0} & \textbf{70.2} \\
        Claude-3-7-Sonnet & 79.5 & 67.2 & 69.5\\
        \multicod{}~+ Claude-3-5-Sonnet & 79.2 & 67.3 & 69.5 \\
        \multicod{}~+ GPT-o1 & 79.0 & 66.8 & 69.0 \\
         \multicod{}~+ Qwen2.5-Coder-32B (Ours) & 78.9 & 66.5 & 68.9\\
        Qwen2.5-Coder-32B (Standard) & 77.1 & 64.0 & 67.8\\
        \bottomrule
    \end{tabular}}
    \label{tab:defects4j}
\end{table}

While the improvements on Defects4J are more modest than on other benchmarks, they still represent significant progress. The \multicod{}~approach with Qwen2.5-Coder-32B achieves 98.5\% of \multicod{}~+ Claude-3-7-Sonnet's Compilation Rate, 97.8\% of its Pass@1, and 98.2\% of its BLEU score. More importantly, it substantially outperforms standard Qwen2.5-Coder-32B (improving CR by 1.8\%, Pass@1 by 2.5\%, and BLEU by 1.1\%) and narrows the gap with Claude-3-7-Sonnet to just 0.6\% on Compilation Rate, 0.7\% on Pass@1, and 0.6\% on BLEU. These results demonstrate that our approach effectively elevates open-source model performance to near closed-source levels even on challenging real-world Java program repair tasks.






\subsection{Case Study: \textsc{Multi-CoD} in Action}
\mpara{Task Context.} We analyze our framework on BigCodeBench/5: generating a dictionary mapping letters to standard deviations of random integer lists.

\mpara{Strategy Generation.} Figure~\ref{fig:multi_cod_example} shows five generated strategies: Dictionary-First, Statistical Function, Generator-Based, Parallel Processing, and Functional Composition, each with distinct approaches to the task.

\mpara{Solution Synthesis.} All generated solutions maintained 100\% CoD adherence while following their strategic guidance—from Dictionary-First's upfront data structure to Functional Composition's modular approach.

\mpara{Feature Analysis \& Selection.} The VADN correctly assigned the highest Q-value (30.20) to the Functional Composition solution, recognizing its superior structure (high function count, no loops) and perfect CoD adherence. This was the only solution that passed all test cases.

\mpara{Insights.} This case study demonstrates \textsc{Multi-CoD}'s effectiveness in generating diverse solutions, maintaining reasoning constraints, and intelligently selecting the optimal candidate, balancing exploration and exploitation while evaluating only one of five solutions.

\begin{figure*}[htbp]
\small
    \centering
    \begin{mdframed}[style=niceframe, frametitle={\textbf{Running Example: \multicod~for Task BigCodeBench/5}}]
    \small
    \textbf{Task}: Generate a function that creates a dictionary where each key is a lowercase letter and each value is the population standard deviation of a random list of integers.
    
    \vspace{0.5em}
    \begin{minipage}[t]{0.98\textwidth}
        \textbf{Step 1: Strategy-Guided Prompt Generation}\\
        \resizebox{\linewidth}{!}{
        \begin{tabular}{p{0.19\textwidth}|p{0.19\textwidth}|p{0.19\textwidth}|p{0.19\textwidth}|p{0.19\textwidth}}
        \toprule
        \textbf{Strategy 1} & \textbf{Strategy 2} & \textbf{Strategy 3} & \textbf{Strategy 4} & \textbf{Strategy 5} \\
        Dictionary-First & Statistical Function & Generator-Based & Parallel Processing & Functional Composition \\
        \midrule
        \textit{Focus}: Building data structure before calculations & \textit{Focus}: Creating reusable statistical components & \textit{Focus}: Using generators for memory efficiency & \textit{Focus}: Optimizing for larger datasets & \textit{Focus}: Breaking down into pure functions \\
        \midrule
        \textit{Instruction}: 
        1. Create empty dict 
        2. Generate random lists 
        3. Add to dict 
        4. Calculate SD
        5. Build result dict & 
        \textit{Instruction}: 
        1. Write SD function 
        2. Test SD function
        3. Generate data
        4. Apply SD
        5. Return results & 
        \textit{Instruction}: 
        1. Create number generator
        2. Generate lists
        3. Calculate SDs
        4. Yield results
        5. Build dict & 
        \textit{Instruction}: 
        1. Split letter groups
        2. Process chunks
        3. Calculate parallel
        4. Merge results
        5. Format output &
        \textit{Instruction}: 
        1. Create random list function
        2. Create SD calculator
        3. Map letters
        4. Compose functions
        5. Return dict \\
        \bottomrule
        \end{tabular}}
    \end{minipage}
    
    \vspace{0.5em}
    \begin{minipage}[t]{0.95\textwidth}
        \textbf{Step 2: CoD-Constrained Solution Synthesis}\\
        \begin{tabular}{p{0.18\textwidth}|p{0.18\textwidth}|p{0.18\textwidth}|p{0.18\textwidth}|p{0.18\textwidth}}
        \toprule
        \textbf{Strategy 1} & \textbf{Strategy 2} & \textbf{Strategy 3} & \textbf{Strategy 4} & \textbf{Strategy 5} \\
        \midrule
        \begin{lstlisting}[basicstyle=\tiny\ttfamily]
import random
import math

def task_func(LETTERS=[chr(i) for i in range(97, 123)]):
    # Initialize dictionary
    number_dict = {}
    
    # Generate random lists
       ...
        \end{lstlisting} & 
        \begin{lstlisting}[basicstyle=\tiny\ttfamily]
import random
import math

def task_func(LETTERS=[chr(i) for i in range(97, 123)]):
    def calculate_mean(numbers):
        return sum(numbers) / len(numbers)
        ...
        \end{lstlisting} & 
        \begin{lstlisting}[basicstyle=\tiny\ttfamily]
import random
import math

def task_func(LETTERS=[chr(i) for i in range(97, 123)]):
    def generate_random_list():
        ...
        \end{lstlisting} & 
        \begin{lstlisting}[basicstyle=\tiny\ttfamily]
import random
import math

def task_func(LETTERS=[chr(i) for i in range(97, 123)]):
    # Initialize result dictionary
    result_dict = {}
 ...
        \end{lstlisting} & 
        \begin{lstlisting}[basicstyle=\tiny\ttfamily]
import random
import math

def task_func(LETTERS=[chr(i) for i in range(97, 123)]):
    def generate_random_list():
...
        \end{lstlisting} \\
        \bottomrule
        \end{tabular}
    \end{minipage}
    
    \vspace{0.5em}
\begin{minipage}[t]{0.98\textwidth}
    \textbf{Step 2 (continued): CoD Steps}\\
    \resizebox{\linewidth}{!}{
    \begin{tabular}{p{0.19\textwidth}|p{0.19\textwidth}|p{0.19\textwidth}|p{0.19\textwidth}|p{0.19\textwidth}}
    \toprule
    \textbf{Strategy 1 Steps} & \textbf{Strategy 2 Steps} & \textbf{Strategy 3 Steps} & \textbf{Strategy 4 Steps} & \textbf{Strategy 5 Steps} \\
    \midrule
    1. Initialize empty dictionary; 2. Generate random list length; 3. Create random integer lists; 4. Store lists in dictionary; 5. Calculate mean for each list; 6. Calculate squared differences; 7. Calculate population standard deviation; 8. Create final dictionary & 
    1. Define population SD function; 2. Create mean calculation helper; 3. Test SD with sample data; 4. Generate random number dictionary; 5. Calculate SD for each list; 6. Return final SD dictionary & 
    1. Create random number generator function; 2. Generate random list length; 3. Yield random integer lists; 4. Calculate mean of list; 5. Calculate squared differences generator; 6. Calculate population standard deviation; 7. Build final dictionary result & 
    1. Initialize empty result dictionary; 2. Generate random lists per letter; 3. Calculate mean for each list; 4. Calculate squared differences from mean; 5. Calculate population standard deviation; 6. Store results in dictionary; 7. Return formatted dictionary & 
    1. Define random list generator function; 2. Create standard deviation calculator; 3. Generate random lists dictionary; 4. Calculate mean helper function; 5. Map letters to random lists; 6. Calculate SD for each list; 7. Create final dictionary mapping \\
    \midrule
    Adherence: 100\%, Avg. Words: 3.75 & Adherence: 100\%, Avg. Words: 4.33 & Adherence: 100\%, Avg. Words: 4.14 & Adherence: 100\%, Avg. Words: 4.29 & Adherence: 100\%, Avg. Words: 4.43 \\
    \bottomrule
    \end{tabular}}
\end{minipage}
    
\vspace{0.5em}
\begin{minipage}[t]{0.98\textwidth}
    \begin{minipage}[t]{0.58\textwidth}
        \textbf{Step 3: Feature Extraction (Strategy 5)}
        
        \resizebox{\linewidth}{!}{
        \begin{tabular}{l|l|l|l}
        \toprule
        \textbf{Code Complexity} & \textbf{CoD Metrics} & \textbf{Strategy Metadata} & \textbf{Relative Features} \\
        \midrule
        Char count: 1295.0 & Adherence rate: 1.0 & Strategy index: 4 & Char count ratio: 1.0 \\
        Line count: 36.0 & Total draft steps: 7.0 & Temperature: 0.8 & Line count ratio: 1.0 \\
        Function count: 6.0 & Avg words per step: 4.43 & Time-focused: 0.0 & Draft steps ratio: 1.0 \\
        Loop count: 0.0 & Min words per step: 2.58 & Space-focused: 0.0 & Rank by length: 0.25 \\
        Conditional count: 1.0 & Max words per step: 4.43 & & Rank by adherence: 0.0 \\
        Comment count: 6.0 & Std dev words: 0.24 & & Is shortest: 0.0 \\
        \bottomrule
        \end{tabular}}
    \end{minipage}
    \hfill
    \begin{minipage}[t]{0.39\textwidth}
        \textbf{Step 4: VADN Solution Selection}
        
        \resizebox{\linewidth}{!}{
        \begin{tabular}{l|c|c|c}
        \toprule
        \textbf{Strategy} & \textbf{Q-Value} & \textbf{Pass} & \textbf{Reward} \\
        \midrule
        Dictionary-First & 5.17 & No & - \\
        Statistical Function & 5.78 & No & - \\
        Generator-Based & 24.14 & No & - \\
        Parallel Processing & -2.84 & No & - \\
        \rowcolor{Gray} \textbf{Functional Composition} & \textbf{30.20} & \textbf{Yes} & \textbf{+1.7} \\
        \bottomrule
        \end{tabular}}
    \end{minipage}
\end{minipage}
    \vspace{0.5em}
    \begin{minipage}[t]{0.95\textwidth}
        \textbf{Step 5: Final Results}
        
        The VADN successfully selected the Functional Composition Method (Strategy 5), which was also the only solution that passed the test cases. The reward of +1.7 includes the base reward for a correct solution (+1.0), a bonus for selecting the first passing solution (+0.5), and a bonus for the solution with best CoD adherence (+0.2).
        
        The final solution uses pure functions with clear separation of concerns:
        \ding{182} A function to generate random lists;
        \ding{183} A helper function to calculate means;
        \ding{184} A standard deviation calculator;
        \ding{185} Functions to create and transform dictionaries
        
        The functional approach was preferred by VADN due to its well-structured code (function count: 6), complete adherence to CoD constraints, and appropriate complexity metrics. This example demonstrates how VADN effectively identifies the optimal solution from multiple candidates by considering multiple feature dimensions.
    \end{minipage}
    \end{mdframed}
    \caption{Complete execution of the \multicod~framework on a BigCodeBench task, showing the progression from strategy generation through solution synthesis to final selection via VADN.}
    \label{fig:multi_cod_example}
\end{figure*}

\subsection{Token Consumption and Generation Time Analysis}

Here, we analyze the computational efficiency, focusing on generation time and output token consumption across benchmarks. Time and token costs are computed as described in Section~\ref{sec:token-time-gen-compute}.

\noindent
\textbf{Generation Time Analysis.}
In Table~\ref{tab:generation_time}, we present the average generation time per problem for each reasoning strategy.

\begin{table}[t]
\centering
\caption{Average generation time (seconds) by reasoning strategy}
\label{tab:generation_time}
\resizebox{\linewidth}{!}{
\begin{tabular}{l|r|r|r|r|r}
\toprule
\textbf{Benchmark} & \textbf{Standard} & \textbf{CoT} & \textbf{Chain-of-Draft} & \textbf{Multi-CoD (best)} & \textbf{Multi-CoD (best)/CoT Ratio} \\
\midrule
BigCodeBench & 5.2 & 12.8 & 6.4 & 6.8 & 0.53$\times$ \\
MBPP & 3.7 & 9.2 & 4.8 & 5.1 & 0.55$\times$ \\
SWE-bench & 14.6 & 35.9 & 18.2 & 19.3 & 0.54$\times$ \\
Defects4J & 17.3 & 42.1 & 20.9 & 22.4 & 0.53$\times$ \\
\bottomrule
\end{tabular}}
\end{table}

The time analysis reveals that Chain-of-Draft consistently generates solutions in approximately half the time required by Chain-of-Thought across all benchmarks, with Multi-CoD (best) requiring only a modest 5-6\% overhead for selection. Despite this slight increase, Multi-CoD (best) still maintains a significant time advantage over CoT, requiring only 53-55\% of CoT's generation time across all benchmarks. Complex software engineering tasks (SWE-bench and Defects4J) demand significantly longer generation times than simpler tasks (MBPP and BigCodeBench) across all strategies, reflecting the greater complexity of working with larger codebases.

\noindent
\textbf{Analyzing Output Token Consumption.}
Table~\ref{tab:full_token_consumption} shows the average output token counts per problem.

\begin{table}[h]
\centering
\caption{Average output token consumption per problem.}
\label{tab:full_token_consumption}
\resizebox{\linewidth}{!}{
\begin{tabular}{lccccc}
\toprule
\textbf{Method} & \textbf{BigCodeBench} & \textbf{MBPP} & \textbf{SWE-bench} & \textbf{Defects4J} \\
\midrule
Strategy Generation & 325 & 280 & 468 & 512 \\
5 CoD Solutions & 2,590 & 1,980 & 8,167 & 9,508 \\
\textbf{Total \multicod{}}(all) & \textbf{2,915} & \textbf{2,260} & \textbf{8,635} & \textbf{10,020} \\
\textbf{Total \multicod{}}(best) & \textbf{596} & \textbf{471} & \textbf{1,952} & \textbf{2,245}\\
\midrule
Standard Prompting & 410 & 326 & 1,243 & 1,572 \\
CoD (Single Solution) & 583 & 452 & 1,927 & 2,204 \\
CoT (Single Solution) & 986 & 765 & 3,894 & 4,618 \\
\bottomrule
\end{tabular}}
\end{table}



While Multi-CoD(all) uses approximately 2.2-3.0× more tokens than CoT and 6.4-7.1× more than standard prompting, the absolute performance gains are substantial, particularly on complex tasks like SWE-bench where Multi-CoD achieves 62.7\% resolution rate compared to 47.6\% for CoT—a 15.1 percentage point improvement.

To further understand efficiency, we focus on \multicod{}(best) which reflects the actual user-billed scenario, and balances performance and token cost.
Our analysis shows that \multicod{}(best)'s token advantage over CoT increases with task complexity: for simpler tasks (BigCodeBench, MBPP), it uses 60-62\% of CoT's tokens, while for complex tasks (SWE-bench, Defects4J), this drops to 49-50\%. 
We observe that Multi-CoD (best) uses only 692 tokens compared to CoT's 2,451 tokens on SWE-bench—a 71.8\% reduction. Importantly, the RL-selected best solution from Multi-CoD requires virtually the same tokens as a single CoD solution (only 1-4\% more), demonstrating that Multi-CoD's performance benefits come with minimal token overhead.



\begin{figure}[t]
    \centering
    \includegraphics[width=\linewidth]{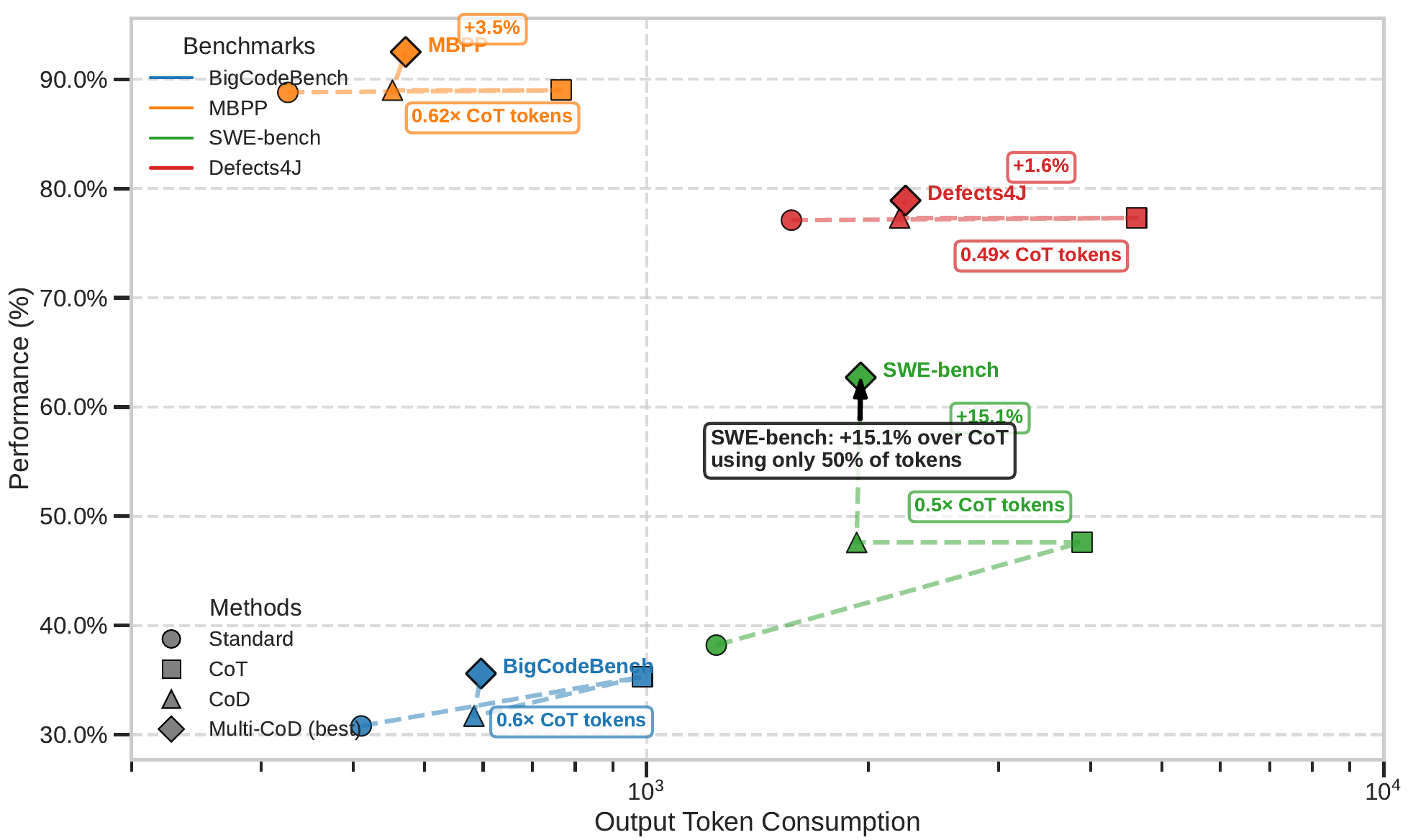}
    \caption{Performance vs. output token consumption across benchmarks. Multi-CoD (best) uses approximately half the tokens of Chain-of-Thought while delivering substantially higher performance, particularly on complex tasks like SWE-bench where it improves resolution rate by 24.5 percentage points over standard approaches and 15.1 percentage points over CoT.}
    \label{fig:performance_vs_tokens}
\end{figure}

\noindent
\textbf{Performance-Efficiency Tradeoff.}
Figure~\ref{fig:performance_vs_tokens} shows the performance and efficiency tradeoff.
Our token efficiency analysis shows that Multi-CoD (best) consistently achieves better performance than CoT while using only 49-62\% of the tokens, with savings primarily from concise reasoning steps. On SWE-bench, Multi-CoD improves resolution rate by 24.5 percentage points over standard approaches and 15.1 percentage points over CoT, while using only 50\% of the tokens required by CoT. This demonstrates exceptional efficiency on complex tasks where reasoning verbosity in CoT becomes excessive. For simpler tasks like BigCodeBench and MBPP, Multi-CoD (best) matches or exceeds CoT's performance while still maintaining a significant token advantage, using only 60-62\% of CoT's tokens.

\subsection{Ranker Comparison on MBPP Benchmark}
\label{sec:selector_comparison}

While ranking methods may appear to be a natural substitute for the selection component (VADN) in \multicod{}, they are fundamentally orthogonal to our approach. Existing rankers such as CodeReviewer~\cite{zhang2023coder}, Lever~\cite{ni2023lever}, and CodeRanker~\cite{inala2022fault} primarily operate on a fixed pool of one-shot completions, ranking candidates based on execution feedback or error traces. These methods typically train a supervised model to predict pass/fail outcomes or fault types, as exemplified by RankEF~\cite{sun2024sifting}, the current state-of-the-art. In contrast, \multicod{} does not rely on execution feedback or static ranking; instead, it actively generates diverse reasoning strategies and learns a policy to select the most promising candidate using interpretable features, making it complementary rather than interchangeable with rankers.

For completeness, we compare \multicod{} against state-of-the-art code ranking methods on the MBPP benchmark. Specifically, we use RankEF~\cite{sun2024sifting} and CodeRanker~\cite{inala2022fault}, which represent the strongest ranking-based approaches reported in prior work. Additionally, we report results for a Random Selection baseline to contextualize the effectiveness of learned selection strategies.

\subsubsection{Experimental Setup}

For fair comparison, we modify our pipeline to evaluate different selection mechanisms while keeping generation consistent. All approaches generate $k=5$ candidate solutions using our CoD-constrained strategy with identical temperature settings. The key difference lies in the selection mechanism: Multi-CoD uses our trained VADN, while comparison methods use their respective ranking models.

\subsubsection{Performance Comparison}

\begin{table}[h]
\centering
\caption{Selector Performance on MBPP with CoD-Generated Candidates}
\label{tab:selector_mbpp}
\begin{tabular}{lccc}
\toprule
\textbf{Selection Method} & \textbf{Pass@1} & \textbf{Pass@2} & \textbf{Pass@5} \\
\midrule
Random Selection & 88.8 & 89.5 & 90.2 \\
CodeRanker & 90.1 & 91.2 & 91.5 \\
RankEF & 91.8 & 92.4 & 93.1 \\
Multi-CoD (VADN) & \textbf{92.5} & \textbf{93.2} & \textbf{93.8} \\
\bottomrule
\end{tabular}
\end{table}

In general, we observe that RankEF outperforms CodeRanker and Random selection.
We summarize three key findings from our comparison:
(i) \textbf{Selection Quality.} Multi-CoD's VADN selector outperforms RankEF by 0.7\% on Pass@1, despite RankEF using execution feedback and being specifically designed for code ranking. (ii) \textbf{Training Efficiency:} While RankEF requires fine-tuning on code generation datasets with execution feedback, our VADN demonstrates strong generalization to MBPP despite being trained on a different dataset entirely. (iii) \textbf{Inference Speed.} Multi-CoD selection requires only feature extraction and a forward pass through VADN (less than 0.1 seconds), while RankEF needs to generate execution feedback representations for each candidate.

The VADN’s ability to outperform RankEF without requiring execution feedback or extensive fine-tuning demonstrates the value of our interpretable feature-based approach. For practitioners, the choice between single-shot and multi-candidate approaches depends on the specific tradeoff between token cost and solution quality required by their application.

\subsection{Practical Considerations for \multicod{}}
Although \textsc{Multi-CoD} introduces additional token overhead, several factors mitigate this overhead in practice:
\begin{itemize}[leftmargin=*]
\item \textbf{Parallel Generation:} Candidate solutions can be generated concurrently, reducing generation time to that of a single generation plus selection overhead.
\item \textbf{Caching Strategy:} For frequently encountered problem patterns, strategy templates can be cached and reused, eliminating strategy generation tokens.
\item \textbf{Adaptive k-Selection:} In production, $k$ can be dynamically adjusted based on problem difficulty, using fewer candidates for simpler problems while maintaining $k=5$ for complex tasks.
\end{itemize}


For applications where correctness is paramount, such as production code generation or critical bug fixes, these performance improvements justify the additional token cost. However, for simpler tasks or token-constrained environments, single-shot approaches may be more appropriate. The framework's modular design allows practitioners to adjust $k$ based on their specific performance-cost requirements.

\section{Related Work}

Reasoning and code generation have attracted growing attention with the rise of large language models (LLMs). Yang et al.~\cite{yang2025code} distinguish between \textit{code-enhanced reasoning}, where code supports general reasoning (e.g., PaL~\cite{gao2023pal}, PoT~\cite{chen2022program}), and \textit{reasoning-enhanced code intelligence}, where structured reasoning improves code generation.

Our work falls into the latter. Structured prompting methods like Chain-of-Thought (CoT)~\cite{wei2022chain} enhance code synthesis via explicit reasoning steps~\cite{huang2023codecot, li2025structured, yin2024thinkrepair}. Self-consistency sampling improves CoT by selecting the most consistent path~\cite{wang2022self}. More advanced methods, such as Tree-of-Thought~\cite{yao2023tree} and Graph-of-Thought~\cite{besta2024graph}, explore reasoning as structured search. Other extensions include uncertainty-aware CoT~\cite{zhu2025uncertainty}, curriculum-guided reasoning via CRPE~\cite{gui2025crpe}, and steering between textual and executable reasoning~\cite{chen2024steering}. Symbolic planning approaches like Self-Planning~\cite{jiang2024self}, CodeChain~\cite{lecodechain}, and CodePlan~\cite{bairi2024codeplan} introduce multi-step decomposition but often suffer from verbosity and inefficiency.

Efficient LLM deployment requires minimizing computational cost, particularly for reasoning-heavy tasks. Speculative decoding~\cite{chen2023accelerating, leviathan2023fast} and FlashAttention~\cite{dao2022flashattention} accelerate inference, while methods like TELL~\cite{zeng2024tell} reduce training data needs. Multi-token prediction improves speed and accuracy~\cite{gloeckle2024better}, and token pruning~\cite{keith2024optimizing}, attention-guided dropping~\cite{arif2025hired}, and activation filtering (Collider)~\cite{chai2025enhancing} further optimize efficiency. Still, most reasoning-based techniques trade efficiency for accuracy.
Chain-of-Draft (CoD) prompting~\cite{xu2025chain} offers a concise alternative to verbose CoT by generating minimal intermediate drafts. Yet, its stochastic outputs pose challenges in identifying the best solution.

Reinforcement Learning has shown promise in guiding LLMs using feedback signals. CodeRL~\cite{le2022coderl} applies deep RL with test-based rewards. PPOCoder~\cite{shojaee2023execution} improves stability, while RLTF~\cite{liu2023rltf} uses fine-grained feedback. StepCoder~\cite{dou2024stepcoder} incorporates compiler signals for incremental optimization. Process-supervised RL~\cite{ye2025process} and systems like FALCON~\cite{li2024falcon} further improve generation through structured memory. 

\section{Conclusion}
In this work, we introduced \multicod, a reinforcement learning-guided framework that enhances the Chain-of-Draft (CoD) prompting paradigm by modeling solution selection as a contextual bandit problem to identify the most promising candidate from diverse CoD-generated solutions. \multicod~performs on par with, and in many cases outperforms existing baselines while significantly reducing token charges (often by over 50\%) per solution, and improving LLM response quality. 
Finally, our approach substantially narrows the performance gap between open and closed-source models for code generation while paving the way for more sustainable and scalable reasoning frameworks.

\noindent{\bf Open Science:} All artefacts are available at:
\begin{center}
{\url{https://anonymous.4open.science/r/MultiCoD}}
\end{center}

\newpage

\newpage
\balance
\bibliographystyle{IEEEtran}
\bibliography{main}

\end{document}